\documentclass[column,showpacs,preprintnumbers,amsmath,amssymb]{revtex4}
\preprint{}
\usepackage{graphicx}
\usepackage{dcolumn}
\usepackage{bm}
\usepackage{mathrsfs}
\begin{document}
\title{Optimal Quantum Thermometry by Dephasing}
\author{Dong  Xie}
\email{xiedong@mail.ustc.edu.cn}
\affiliation{Faculty of Science, Guilin University of Aerospace Technology, Guilin, Guangxi, P.R. China.}

\author{Chunling Xu}
\affiliation{Faculty of Science, Guilin University of Aerospace Technology, Guilin, Guangxi, P.R. China.}

\author{An Min Wang}
\affiliation{Department of Modern Physics , University of Science and Technology of China, Hefei, Anhui, China.}

\begin{abstract}
Decoherence often happens in the quantum world. We try to utilize quantum dephasing to build an optimal thermometry.
By calculating the Cram$\acute{e}$r-Rao bound, we prove that the Ramsey measurement is the optimal way to measure the temperature for uncorrelated particles. Using the optimal measurement, the metrological equivalence of product and maximally entangled states of initial quantum probes that always holds. However, using Ramsey measurement,  the metrological equivalence only holds in special situation. Contrary to frequency estimation, the quantum limit can be surpassed under the case $\nu<1$. For the general Zeno regime($\nu=2$), uncorrelated product states are the optimal choose in typical Ramsey spectroscopy set-up. In order to surpass the standard scaling, we propose to change the interaction strength with time. Finally, we investigate other environmental influences on the measurement precision of temperature. Base on it, we define a new way to measure non-Markovian effect.
\end{abstract}

\maketitle

\section{Introduction}
With the development of quantum technology, utilizing quantum probes to measure the temperature of a sample accurately is becoming a central task in modern physics\cite{lab1}. Creating precise thermometers is very meaningful in understanding thermodynamics. And it will bring breakthroughs in other fields, such as medicine, biology, and material science.

Recently, Correa et.al\cite{lab2} investigate the fundamental limitations on temperature estimation with an individual quantum probe.
As a result, whether the quantum probe is thermalized completely or not, quantum coherence in the initial state of the probe is not directly linked to the overall maximization of the precision. However, in some simple thermometric tasks, coherence and entanglement play an important role\cite{lab3,lab4}. In the reference \cite{lab5}, the authors consider that thermometry may be mapped on to the problem of phase estimation, it follows that the scaling of the precision of a thermometer may in principle be improved to Heisenberg limit with entanglement states. However, the constraints on the bath and the interaction Hamiltonian in the toy model are  difficult to be satisfied.

Due to the interaction with other environments, quantum decoherence\cite{lab6} often happens. So we consider that quantum probes suffer from dephasing due to thermal contact with a sample. Two kinds of dephasing process are considered: Markovian dephasing and non-Markovian dephasing. The dephasing rate can be described by the general power law form, $\gamma(t)=\alpha(T)t^\nu$, where the temperature $T$ is encoded in the factor $\alpha(T)$. The case of $\nu = 1$ corresponds to the Markovian case. The case of time-inhomogeneous ($\nu\neq 1$ ) corresponds to non-Markovian case.
We obtain the maximal quantum Fisher information by variational approach, and find that Ramsey measurement is the optimal way to measure the temperature by dephasing dynamics for initial probes in product state. Using Ramsey measurement: we show that the metrological equivalence of product and maximally entangled states of initial quantum probes holds for $\nu\leq1$ under special conditions; for non-Markovian dephasing case, the quantum limit can be surpassed only under the case $\nu<1$; for the general non-Markovian dephasing(the Zeno regime, $\nu=2$), uncorrelated probes are in favour of the estimation of temperature; correlated state never performs better than uncorrelated state. Using the optimal measurement: the metrological equivalence of product and maximally entangled states of initial quantum probes that always holds. When the optimal condition can not be satisfied, we find that maximally entangled state can perform better than product state for $\nu<1$.
 In order to surpass the quantum limit, we propose to change the interaction strength with time to obtain $\nu<1$. Finally, we study that the quantum probes suffer from other environments besides the sample to be tested. And we define a new way to quantify non-Markovianity in terms of the nonmonotonicity of the quantum Fisher information about temperature.

The rest of this article is arranged as following. In section II, we briefly introduce the quantum Fisher information and the variational approach for dealing with mixed state. In section III, the optimal precision of temperature by Ramsey measurement and optimal measurement are obtained. We prove that Ramsey measurement is the optimal way. Then, for Markovian and non-Markovian dephasing process, we compare the function of product and maximally entangled states of initial quantum probes on the metrological  precision.
In Section IV, there are other environments which can influence the measurement of temperature. A new way to quantify non-Markovian effect from the other environments is proposed in Section V. A conclusion and outlook are presented in Section VI.
\section{review of quantum Fisher information and variational approach}
The famous Cram$\acute{e}$r-Rao bound\cite{lab7,lab8,lab9} offers a very good parameter estimation under the constraints of quantum physics:
\begin{eqnarray}
(\delta x)^2\geq\frac{1}{N\mathcal{F}_Q[\hat{\rho}_S(x)]},
\end{eqnarray}
where $N=\tau/t$ represents total times of experiments given by the fixed total time $\tau$.  $\mathcal{F}_Q[\hat{\rho}_S(x)]$ denotes quantum Fisher information, which can be generalized from classical Fisher information. The classical Fisher information is defined by
\begin{equation}
f(x)=\sum_k p_k(x)[d\ln[p_k(x)]/dx]^2,
\end{equation}
where $p_k(x)$ is the probability of obtaining the set of experimental results $k$ for the parameter value $x$. Furthermore,
the QFI is given by the maximum of the Fisher information over all measurement strategies allowed by quantum physics:
\begin{equation}
\mathcal {F}_{Q}[\hat{\rho}(x)]=\max_{\{\hat{E}_k\}}f[\hat{\rho}(x);\{\hat{E}_k\}],
\end{equation}
where positive operator-valued measure $\{\hat{E}_k\}$ represents a specific measurement device.

If the probe state is pure, $\hat{\rho}_S(x)=|\psi(x)\rangle\langle\psi(x)|$, the corresponding expression of QFI is
\begin{equation}
\mathcal {F}_{Q}[\hat{\rho}(x)]=4[\frac{d\langle\psi(x)|}{dx}\frac{d|\psi(x)\rangle}{dx}-|\frac{d\langle\psi(x)|}{dx}|\psi(x)\rangle|^2].
\end{equation}

When the probe state is mixed state, the question becomes complex.  It is always possible to enlarge
the size of the original Hilbert space $S$ and build a pure state $|\Phi_{S,E}(x)\rangle\langle\Phi_{S,E}(x)|$ in the enlarged space $S + E$ that fulfills the condition $\textmd{Tr}_E[|\Phi_{S,E}(x)\rangle\langle\Phi_{S,E}(x)|]=\hat{\rho}_S(x)$. Here the state $|\Phi_{S,E}(x)\rangle$ is a purification of mixed state $\hat{\rho}_S(x)$.  Although there are many purification of mixed state, any two purifications of mixed state $\hat{\rho}_S(x)$ are always connected by a unitary operators $\hat{u}_E(x)$.

When a system and an imaginary environment are monitored together, the information acquired about the unknown parameter can not be
smaller than the information obtained when only the system is measured. Thus, the QFI can be obtained as follows:
\begin{equation}
 \mathcal{{F}}_{Q}[\hat{\rho}_S(x)]= \min_{|\Phi_{S,E}(x)\rangle}\mathcal{F}_Q[\hat{\rho}_{S,E}(x)].
\end{equation}
Utilizing the unitary operator, the QFI can be described by
\begin{equation}
 {F}_{Q}[\hat{\rho}_S(x)]= \min_{\hat{u}_E(x)}\mathcal{F}_Q[\hat{\rho}_{S,E}(x)].
\end{equation}
Then, the total Hamiltonian $\hat{H}_{S,E}$ and imaginary environmental Hamiltonian $\hat{h}_E(x)$ can be denoted as
\begin{equation}
\hat{h}_E(x)=i\frac{d\hat{u}_E^\dagger(x)}{dx}\hat{u}_E(x),\\
i\frac{d|\Phi_{S,E}(x)\rangle}{dx}=\hat{H}_{S,E}(x)|\Phi_{S,E}(x)\rangle.
\end{equation}
Substituting the above definitions and Eq.(6) into Eq.(4), the QFI can be expressed by
\begin{equation}
\mathcal {F}_{Q}[\hat{\rho}_S(\phi)]=\min_{\hat{h}_E(\phi)}4\langle[ \hat{\mathcal{H}}(\phi)-\langle\hat{\mathcal{H}}(\phi)\rangle]^2\rangle_\Phi,
\end{equation}
in which,
\begin{equation}
\hat{\mathcal{H}}(\phi)=\hat{H}_{S,E}(\phi)-\hat{h}_E(\phi).\nonumber
\end{equation}
 The optimum Hermitian operator $\hat{h}^{\textmd{opt}}_E(\phi)$) should satisfy the following equation:
 \begin{equation}
\hat{h}_E^{(opt)}\hat{\rho}_E(x)+\hat{\rho}_E(x)\hat{h}_E^{(opt)}=i\textmd{Tr}_S[\frac{d|\Phi_{S,E}\rangle}{dx}\langle\Phi_{S,E}|-|\Phi_{S,E}\rangle\frac{d\langle\Phi_{S,E}|}{dx}].
\end{equation}
 It is difficult to solve the above equation. One can use variation approach\cite{lab10}: first step, guessing the approximation for $\hat{h}_E^{(opt)}$, which depends on the variational parameter; second step, obtain the optimal Hermitian operator $\hat{h}^{\textmd{opt}}_E(\phi)$) by taking a derivative with respect to the variational parameter. For solving question, some symmetry can be used to guess the form of  $\hat{h}_E^{(opt)}$.

\section{optimal precision of temperature}
We consider a global system composed of n particles. The Hamiltonian of each particle is described by $\frac{w_0}{2}Z$ ( $\hbar = 1$ throughout), where the eigenvectors of the Pauli operator $Z=|1\rangle\langle1|-|0\rangle\langle0|$ are denoted by ($|0\rangle, |1\rangle$) . The $n$ particles suffer from the correspond $n$ uncorrelated parts of a sample, which induce pure dephasing.
The time evolution of the reduced density matrix of the system (for one particle) is denoted by
\begin{equation}
\rho_{ii}(t)=\rho_{ii}(0),
\end{equation}
\begin{equation}
\rho_{01}(t)=\rho_{01}(0)e^{-2\gamma(t)},
\end{equation}
where $i=0,1$.
The temperature $T$ of sample can be encoded in the dephasing factor $\gamma(t)$. Without loss of generality, we consider the dephasing factor has a general power law dependence on time: $\gamma(t)=\alpha(T)t^\nu$. It is due to that the result only depends on the power $\nu$\cite{lab11,lab12}. When $\nu=1$, it corresponds to Markovian dephasing; otherwise, it is non-Markovian case\cite{lab11}. Here we emphasize that non-Makovian dephasing is not non-Makovian effect (non-Markovianity). Non-Markovian effect means that the lost information can come back to system from environment\cite{lab13}.

Firstly, we employ a typical Ramsey spectroscopy set-up\cite{lab14} to measure the $n$ particles.
When the initial state of probes is uncorrelated state $(|0\rangle+|1\rangle)^{\otimes n}$, the resulting single
particle signal is given by
\begin{equation}
p_0=1/2(1+\cos(wt)\exp[-\gamma(t)]), p_1=1/2(1-\cos(wt)\exp[-\gamma(t)]).
\end{equation}
Substituting the above result into Eq.(2), we can obtain the total Fisher information, which is described by
\begin{equation}
\mathcal{F}(T)=n \cos^2(wt)\frac{\exp[-2\gamma(t)](\frac{\partial\gamma(t)}{\partial T})^2 }{1-\cos^2(wt)\exp[-2\gamma(t)]}
\end{equation}
To get the maximal Fisher information, it need that $wt=N\pi$, for $N=0,1,2\cdot\cdot\cdot$.
Using the general power law form $\alpha(T)t^\nu$, we can get the temperature uncertainty
\begin{equation}
\delta T^2|u=\frac{1-\exp[-2\alpha(T)t^\nu]}{n\tau \exp[-2\alpha(T)t^\nu]t^{2\nu-1}(\frac{\partial\alpha(T)}{\partial T})^2}.
\end{equation}
When the dephasing process is Makovian case ($\nu=1$), the optimal precision for uncorrelated initial state is given by
\begin{equation}
\delta T|_u=\sqrt{\frac{2\alpha(T)}{n \tau(\frac{\partial\alpha(T)}{\partial T})^2}},
\end{equation}
where the optimal interrogation time $t\longrightarrow0$.

For an initial preparation of $n$ particles in a maximally entangled state $|0\rangle^{\otimes n}+|1\rangle^{\otimes n}$,  we can make a similar calculation, leading to the result
\begin{equation}
\delta T^2|_e=\frac{1-\exp[-2n\alpha(T)t^\nu]}{n^2\tau \exp[-2n\alpha(T)t^\nu]t^{2\nu-1}(\frac{\partial\alpha(T)}{\partial T})^2}.
\end{equation}

For non-Markovian dephasing case ($\nu\neq1$), we find that for $\nu<1$, the optimal precision $\delta T|_u=\delta T|_e=0<1/n^2$ when the interrogation time $t\longrightarrow0$. It means that the scaling surpass the Heisenberg resolution. Obviously, it is impossible for linear system. The reason comes from the power $\nu<1$.
At short time $t$, the dephasing factor $\gamma(t)\propto t^2$\cite{lab11,lab12,lab15,lab16} must appear. Namely,  the power $\nu<1$ can not appear at short time.

 We suppose that the power $\nu\neq2$ can appear at time $t>t_{cha}$, where $t_{cha}$ is the characteristic time for $\gamma(t)\propto t^2$. Then,  we can obtain the metrological result from product and maximally entangled states of initial quantum probes:
 \[
\textmd{for}\   t_{cha}^\nu\ll1/(n\alpha(T)), n\gg1\Rightarrow  \nu>1/2\left\{
\begin{array}{ll}
  \delta T|_u=\delta T|_e=\sqrt{\frac{2\alpha(T)t_{cha}^{(1-\nu)}}{n \tau(\frac{\partial\alpha(T)}{\partial T})^2}}, &\textmd{for} \  1/2<\nu\leq1;\\
  \delta T|_u\approx\sqrt{\frac{1}{n \tau(\frac{\partial\alpha(T)}{\partial T})^2}[2(2-1/\nu)\alpha(T)]^{(2-1/\nu)}{e^{-(2-1/\nu)}}}, &\textmd{for}\ \nu>1;\\
\delta T|_e\approx\sqrt{\frac{1}{n^2 \tau(\frac{\partial\alpha(T)}{\partial T})^2}[2(2-1/\nu)n\alpha(T)]^{(2-1/\nu)}{e^{-(2-1/\nu)}}}, &\textmd{for}\ \nu>1;
  \end{array}
  \right.
  \]\begin{equation}
 \end{equation}
   \[
\textmd{for}\   1/(\alpha(T))\gg t_{cha}^\nu\gg1/(n\alpha(T)), n\gg1\Rightarrow  \left\{
\begin{array}{ll}

  \delta T|_u\approx\sqrt{\frac{2\alpha(T)t_{cha}^{(1-\nu)}}{n \tau(\frac{\partial\alpha(T)}{\partial T})^2}}, &\textmd{for}\ 1\geq\nu>0;\\
  \delta T|_u\approx\sqrt{\frac{1}{n \tau(\frac{\partial\alpha(T)}{\partial T})^2}[2(2-1/\nu)\alpha(T)]^{(2-1/\nu)}{e^{-(2-1/\nu)}}}, &\textmd{for}\ \nu>1;\\
\delta T|_e\approx\sqrt{\frac{1}{n^2 \tau(\frac{\partial\alpha(T)}{\partial T})^2}[3n\alpha(T)]^{(3/2)}{e^{-(3/2)}}}, &\textmd{for}\ \nu>0;
  \end{array}
  \right.
  \]\begin{equation}
 \end{equation}
  \[
\textmd{for}\   t_{cha}^\nu\gg1/(\alpha(T)), n\gg1\Rightarrow  \left\{
\begin{array}{ll}

  \delta T|_u\approx\sqrt{\frac{1}{n \tau(\frac{\partial\alpha(T)}{\partial T})^2}[3\alpha(T)]^{(3/2)}{e^{-(3/2)}}}, &\textmd{for}\ \nu>0;\\
\delta T|_e\approx\sqrt{\frac{1}{n^2 \tau(\frac{\partial\alpha(T)}{\partial T})^2}[3n\alpha(T)]^{(3/2)}{e^{-(3/2)}}}, &\textmd{for}\ \nu>0;
  \end{array}
  \right.
  \]\begin{equation}
 \end{equation}
  \[
\textmd{for}\   t_{cha}^\nu\approx1/(n\alpha(T)), n\gg1\Rightarrow  \left\{
\begin{array}{ll}

 \delta T|_u=\sqrt{\frac{2\alpha(T)(1/n)^{(1-\nu)}}{n \tau(\frac{\partial\alpha(T)}{\partial T})^2}}, &\textmd{for} \  1>\nu>1/2;\\
 \delta T|_u\approx\sqrt{\frac{1}{n \tau(\frac{\partial\alpha(T)}{\partial T})^2}[2(2-1/\nu)\alpha(T)]^{(2-1/\nu)}{e^{-(2-1/\nu)}}}, &\textmd{for}\ \nu>1;\\
 \delta T|_e\approx\sqrt{\frac{1}{n^2 \tau(\frac{\partial\alpha(T)}{\partial T})^2}[2(2-1/\nu)n\alpha(T)]^{(2-1/\nu)}{e^{-(2-1/\nu)}}}, &\textmd{for}\ 2>\nu>1/2;\\
\delta T|_e\approx\sqrt{\frac{1}{n^2 \tau(\frac{\partial\alpha(T)}{\partial T})^2}[3n\alpha(T)]^{(3/2)}{e^{-(3/2)}}}, &\textmd{for}\ \nu\geq2;
  \end{array}
  \right.
  \]\begin{equation}
 \end{equation}

From above equations, we can see that the maximally entangled state does not offer better resolution of temperature than  the uncorrelated state. In Eq.(17), for the time $ t_{cha}^\nu\ll1/(n\alpha(T))$, the power $\nu>1/2$ is necessary for not surpassing Hensiberg limit. Under this situation, the metrological equivalence of product and maximally entangled states of initial quantum probes holds for $1\geq\gamma>1/2$. And under this condition the quantum limit can be surpassed. For $n\rightarrow\infty$, the $t_{cha}^\nu\ll1/(n\alpha(T))$ does not hold, from Eq.(18) and Eq.(19), we can note that the quantum limit will not be surpassed. For the power $\nu>1$, the metrological equivalence does not hold in any condition, and maximally entangled state obviously reduce the resolution.

So, for improving precision of temperature, the time $t_{cha}$ should be infinitesimal, and the power $\nu<1$ appears after $t_{cha}$. For the general situation with fixed coupling constant, the Markovian dephasing will appear after $t_{cha}$. If one can control the coupling strength between probes and the sample, other powers should appear. For getting the power $\nu<1$, the coupling strength can be proportional to $(t+t_0)^m$, where $0<m<1$, and $t_0$ denotes a constant. And the time $t_{cha}$ is proportional to $t_0$. So, one can reduce the value of  $t_{cha}$ by reducing $t_0$.

Next,  we obtain the optimal precision by optimal measurement. The quantum Fisher information can be obtained by the variational approach in Section II.

At time $t$, a purification of mixed probe state can be described by
\begin{equation}
|\Phi_{S,E}(\phi)\rangle=\prod_{i=1}^ne^{-i\phi t Z_i/2}e^{-i\arccos(\sqrt{P(\gamma t)})Z_iY_i^E}|\psi\rangle|0\rangle^{\otimes n}_E,
\end{equation}
where $|\psi\rangle$ denotes the initial state of probes, and $P(\gamma t)=\frac{1+\exp(-\alpha(T) t^\nu)}{2}$. Operators $Z_i$, $Y_i^E$ represent the Pauli operators of $i$th particle and corresponding environment.
According symmetry, we guess that the operator
\begin{equation}
\hat{h}_E(\phi)=\sum_{i=1}^n\zeta X_i^E+\eta Y_i^E+\delta Z_i^E,
\end{equation}
where $\zeta$, $\eta$ and $\delta $ are variational parameters.

Using Eq.(7) and Eq.(8), by derivation of variational parameters, we can obtain the optimal precision
\begin{equation}
\delta T^2=\frac{1-\exp[-2\alpha(T)t^\nu]}{n\tau \exp[-2\alpha(T)t^\nu]t^{2\nu-1}(\frac{\partial\alpha(T)}{\partial T})^2(1-\langle\sum_{i=1}^{i=n}Z_i/n\rangle)}.
\end{equation}
When the initial probes are in product and maximally entangled state, $\langle\sum_{i=1}^{i=n}Z_i/n\rangle$, the optimal precisions are same. It means that the metrological equivalence holds by using optimal measurement. We find the result by optimal measurement is the same as the the resolution in Eq.(14) by Ramsey measurement. So for uncorrelated particles, Ramsey measurement is the optimal measurement.
\section{Suboptimal precision}
If $wt=N\pi$ in Eq.(13) is not satisfied, the optimal result (Eq.(14)) can not be obtained by Ramsey measurement. In this section, we investigate the optimal precision in unperfect condition $wt=N\pi/4$. We call it suboptimal precision  in comparison to the result in Eq.(14).

The temperature uncertainty is obtained by Ramsey measurement
\begin{eqnarray}
\delta T^2|u=\frac{2-\exp[-2\alpha(T)t^\nu]}{n\tau \exp[-2\alpha(T)t^\nu]t^{2\nu-1}(\frac{\partial\alpha(T)}{\partial T})^2}, \ \textmd{for product state of initial probes};\\
\delta T^2|_e=\frac{2-\exp[-2n\alpha(T)t^\nu]}{n^2\tau \exp[-2n\alpha(T)t^\nu]t^{2\nu-1}(\frac{\partial\alpha(T)}{\partial T})^2}\ \textmd{for maximally entangled  state of initial probes}.
\end{eqnarray}
Like the above section, the power $\nu\neq2$ can appear at time $t>t_{cha}$, where $t_{cha}$ is the characteristic time for $\gamma(t)\propto t^2$. The optimal precision of temperature is given by
 \[
\textmd{for}\   t_{cha}^\nu\leq1/(n\alpha(T)), n\gg1\Rightarrow  \nu>1/2\left\{
\begin{array}{ll}

  \delta T|_u\approx\sqrt{\frac{2}{n \tau(\frac{\partial\alpha(T)}{\partial T})^2}[2(2-1/\nu)\alpha(T)]^{(2-1/\nu)}{e^{-(2-1/\nu)}}} ;\\
\delta T|_e\approx\sqrt{\frac{2}{n^2 \tau(\frac{\partial\alpha(T)}{\partial T})^2}[2(2-1/\nu)n\alpha(T)]^{(2-1/\nu)}{e^{-(2-1/\nu)}}} ;
  \end{array}
  \right.
  \]\begin{equation}
 \end{equation}
 \[
\textmd{for}\   1/(\alpha(T))\gg t_{cha}^\nu\gg1/(n\alpha(T)), n\gg1\Rightarrow  \left\{
\begin{array}{ll}

    \delta T|_u\approx\sqrt{\frac{2}{n \tau(\frac{\partial\alpha(T)}{\partial T})^2}[2(2-1/\nu)\alpha(T)]^{(2-1/\nu)}{e^{-(2-1/\nu)}}}, &\textmd{for}\ \nu>0;\\
\delta T|_e\approx\sqrt{\frac{2}{n^2 \tau(\frac{\partial\alpha(T)}{\partial T})^2}[3n\alpha(T)]^{(3/2)}{e^{-(3/2)}}}, &\textmd{for}\ \nu>0;
  \end{array}
  \right.
  \]\begin{equation}
 \end{equation}
   \[
\textmd{for}\   t_{cha}^\nu\gg1/(\alpha(T)), n\gg1\Rightarrow  \left\{
\begin{array}{ll}

  \delta T|_u\approx\sqrt{\frac{2}{n \tau(\frac{\partial\alpha(T)}{\partial T})^2}[3\alpha(T)]^{(3/2)}{e^{-(3/2)}}}, &\textmd{for}\ \nu>0;\\
\delta T|_e\approx\sqrt{\frac{2}{n^2 \tau(\frac{\partial\alpha(T)}{\partial T})^2}[3n\alpha(T)]^{(3/2)}{e^{-(3/2)}}}, &\textmd{for}\ \nu>0;
  \end{array}
  \right.
  \]\begin{equation}
 \end{equation}
 From Eq.(26), we can find that entangled state can help to improve the precision to surpass the quantum limit for $1>\nu>1/2$, $t_{cha}^\nu\leq1/(n\alpha(T))$. When $\nu=1$, the metrological equivalence of product and maximally entangled states of initial quantum probes that holds. However, for $\nu>1$, the correlated state has opposite effects. For $n\rightarrow\infty$, Eq.(27) and Eq.(28) are applicable. The product state is always better than entangled state.

\section{under external environments}
In this section, we consider that there are external environments: $n$ probe particles suffer from $n$ uncorrelated environments.

Firstly, we consider that environments can induce decoherence with a general dephasing form: $\kappa(t)=\kappa t^{{\nu}'}$. Using Ramsey measurement, the maximal Fisher information can be expressed by
\begin{equation}
\mathcal{F}(T)=n \frac{\exp[-2\gamma(t)-2\kappa(t)](\frac{\partial\gamma(t)}{\partial T})^2 }{1-\exp[-2\gamma(t)-2\kappa(t)]}, \textmd{for product state}
\end{equation}
\begin{equation}
\mathcal{F}(T)=n^2\frac{\exp[-2n\gamma(t)-2n\kappa(t)](\frac{\partial\gamma(t)}{\partial T})^2 }{1-\exp[-2n\gamma(t)-2n\kappa(t]},
\textmd{for maximally entangled state}
\end{equation}
Obviously, decoherence environment can decrease the Fisher information, leading to the reduction of resolution. By analytical calculation, we find that when $\nu\leq\nu '$ and $n\gg1$, the decoherence environments have a small effect on the final effect.
At short time, the dephasing factors are always proportional to $t^2$. So $\nu=\nu'=2$, the decoherence environments plays a more and more smaller effect with the number of particles $n$.

Then, we consider that the environments can induce amplitude damping, which is an usual phenomenon. Tracing over the environments, the quantum operation can be written as
\begin{equation}
\xi_{AD}(\rho_S)=E_0\rho E_0^\dagger+E_1\rho E_1^\dagger,
\end{equation}
in which,
 \[
 E_0= \left[
\begin{array}{ll}
1\ \ 0\\
0\ \ \lambda^{1/2}
  \end{array}
\right ],
  \]
   \[
 E_1= \left[
\begin{array}{ll}
0\    (1-\lambda)^{1/2}\\ 
0\ \ \ \   0
  \end{array}
\right ],
  \]
  where  $\lambda=\exp[-2\int\eta(t)dt]$, and $\eta(t)$ represents the decay rate, such as spontaneous emission.
  Using Ramsey measurement, the maximal Fisher information is given by
  \begin{equation}
\mathcal{F}(T)=n \frac{\exp[-2\gamma(t)-2\int\eta(t)dt](\frac{\partial\gamma(t)}{\partial T})^2 }{1-\exp[-2\gamma(t)-2\int\eta(t)dt]}, \ \textmd{for product state}
\end{equation}
\begin{equation}
\mathcal{F}(T)=n^2\frac{\exp[-2n\gamma(t)-2n\int\eta(t)dt](\frac{\partial\gamma(t)}{\partial T})^2 }{1-\exp[-2n\gamma(t)-2n\int\eta(t)dt]},
\ \textmd{for maximally entangled state}
\end{equation}
We find that in Ramsey set-up the maximal Fisher information for temperature under  amplitude damping environments is the same as the case under decoherence environments.

\section{measuring non-Markovian effect}
If the external environments have memory, there is non-Markovian effect, which is quantified by a lot of ways\cite{lab12,lab17,lab18,lab19,lab20,lab21,lab22,lab23,lab24,lab25,lab26,lab27,lab28}. Among the most important ones, those are based on the deviation of the dynamical maps from divisible CPTP maps\cite{lab17,lab18}and the nonmonotonicity of the trace distance or distinguishability\cite{lab12}.

We quantify the non-Markovian effect as follows:
\begin{equation}
N(\Omega)=\textmd{Max}_{\rho_S(0)}\int_{\mathcal{D}(t)>0}\mathcal{D}(t)dt,
\end{equation}
where $\mathcal{D}(t)=\frac{d}{dt}(\mathcal{F}_E(\rho_S(T))-\mathcal{F}(\rho_S(T)))$. $\mathcal{F}_E(\rho_S(T))$ represents the quantum Fisher information in external environments; $\mathcal{F}(\rho_S(T))$ denotes the case without external environments. The $\rho_S(0)$ denotes the initial state of probe system.
 Here, we need to emphasized that the dephasing factor $\gamma(t)$ and $T$ should be fixed to measure different non-Markovian dynamics $\Omega$. So one can choose proper dephasing process and temperature of sample as a reference. So there are lots of ways to quantify non-Markovian effect.
It is because for any dephasing process and temperature of sample, Eq.(34) always distinguish any non-Markovian effects except
some special cases such as $\frac{\partial\gamma(t)}{\partial T}=0 \ \textmd{and} \ \infty$. As a result, when use Eq.(34) to measure non-Markovianity, one must give the the dephasing factor $\gamma(t)$ and $T$ . At the same time, we must note that the dephasing process should change with the initial state $\rho_S(0)$. For example, when $\rho_S(0)=(|0\rangle+|1\rangle)(\langle0|+\langle1|)$, the dephasing process should be $\rho_S(0)=(|0\rangle\langle0|+|1\rangle\langle1|+\exp[-\gamma(t)]|0\rangle\langle1|+\exp[-\gamma(t)]|1\rangle\langle0|)$; if the initial state becomes $\rho_S(0)=(|+\rangle+|-\rangle)(\langle+|+\langle-|)$,  the dephasing process should be $\rho_S(0)=(|+\rangle\langle-|+|+\rangle\langle-|+\exp[-\gamma(t)]|+\rangle\langle-|+\exp[-\gamma(t)]|+\rangle\langle-|)$.
From the calculation of Eq.(23), we can achieve that the optimal initial state should be like superposition state $|\psi_+\rangle+|\psi_-\rangle$, where $\langle\psi_-|\psi_+\rangle=0$. The corresponding dephasing process should be like $\rho_S(0)=(|\psi_+\rangle\langle\psi_+|+|\psi_-\rangle\langle\psi_-|+\exp[-\gamma(t)]|\psi_+\rangle\langle\psi_-|+\exp[-\gamma(t)]|\psi_-\rangle\langle\psi_+|)$.
So the optimal initial state should be the optimal $|\psi_+\rangle$.

It is easy to verity that when $\mathcal{D}(t)>0$, one can obtain decoherence rate $\frac{\kappa(t)}{dt}<0$ and amplitude rate $\eta(t)<0$ for two levels system. It is the same as the result in reference\cite{lab12}. And it can be generalized to many levels system. Because the way of quantifying non-Markovianity base on the physical essence that when information come back the system again, one can obtain more information about parameter $T$.

Comparing other ways of quantifying non-Markovianity, our way only needs to obtain an optimal  initial state. In reference \cite{lab12} it need to obtain two optimal  initial states. In article \cite{lab28}, it need to obtain the optimal initial state in enlarged space: system+ auxiliary system. So our way is more convenient to obtain the result.
It is interesting to use our way to explore more complex system and environment(beyond the scope of this article).

\section{conclusion and outlook}
We have explore the optimal precision of temperature by quantum dephasing, which often happens in practical quantum
system. Using the typical Ramsey measurement, we find that product state usually performs better than entangled state. Using the optimal measurement way, we use variational approach to demonstrate that the metrological equivalence of product and maximally entangled states of initial quantum probes always holds.  The maximal Fisher information can be much larger than $n$ only in special condition: the characteristic time for dephasing factor $\gamma(t)\propto t^2$ $t_{cha}$. So for enough large value of $n$, the quantum limit will not be surpassed due to that the Zeno effect ($\nu=2$) appears at short time.
For improving the resolution, one must reduce the characteristic time for dephasing factor $\gamma(t)\propto t^2$, and the power $\nu<1$ appears after it.
We also consider imperfect condition: $wt=N\pi/4\neq N\pi$. Under this condition, in the Ramsey measurement set-up, we find that entangled state can perform better than product state at limited range, leading to surpass the quantum limit.
Then we discuss about that there are other environments, which interacts with probes.  For enough particles $n$,  the general decoherence environment plays a very small effect on the final resolution of temperature.  The effect of amplitude damping environments is similar with the case of  decoherence environments.
Finally, base on the Fisher information of temperature, a new way is proposed to quantify non-Markovianity.  Comparing with other measure methods, it is more convenient for calculation.

We believe that this article will help to build an optimal quantum thermometer by quantum dephasing. In this article, we do not concern the interaction among probe particles. The interaction may be a resource for improving the resolution of temperature to surpass quantum limit in general situation, in particular when the number of particles $n$ is enough large. Nonlinear interaction may be help to surpass Hensiberg limit\cite{lab29}.  It is significant to find the optimal interaction form for designing an optimal quantum thermometry.

\section*{Acknowledgement}
This work was supported by the National Natural Science Foundation of China under Grant  No. 11375168.

\end{document}